\documentclass[prl,aps,superscriptaddress,reprint]{revtex4-1}
\usepackage{epsfig}
\usepackage{epsf}
\usepackage{graphicx}
\usepackage{dcolumn}
\usepackage{subfigure}
\usepackage{bm}
\usepackage{tabularx}
\usepackage[dvips]{color}
\usepackage[latin1]{inputenc}
\usepackage{esint}

\def\be{\begin{equation}}
\def\ee{\end{equation}}
\def\ba{\begin{array}}
\def\ea{\end{array}}
\def\bea{\begin{eqnarray}}
\def\eea{\end{eqnarray}}

\def\keff{k_{\mathrm{eff}}}

\newcommand{\Revision}[1]{{\textcolor[named]{Black}{#1}}}

\begin{document}

\title{Quantitative prediction of effective toughness at random heterogeneous interfaces}

\date{\today}

\author{Sylvain Patinet}
\affiliation{
Laboratoire PMMH, ESPCI/CNRS-UMR 7636/Univ. Paris 6 UPMC/Univ. Paris 7 Diderot\\
10 rue Vauquelin, 75231 Paris cedex 05, France}
\affiliation{
Dep. of Materials Science and Engineering, Johns Hopkins University, Baltimore, Maryland 21218, USA}
\author{Damien Vandembroucq}
\affiliation{
Laboratoire PMMH, ESPCI/CNRS-UMR 7636/Univ. Paris 6 UPMC/Univ. Paris 7 Diderot\\
10 rue Vauquelin, 75231 Paris cedex 05, France}
\affiliation{ The Niels Bohr Institut,  Blegdamsvej 17, DK-2100 Copenhagen, Denmark }
\author{St\'ephane Roux}
\affiliation{ LMT-Cachan, ENS de Cachan/CNRS-UMR 8535/Univ. Paris 6 UPMC/PRES
    UniverSud Paris\\
61 Avenue du Pr\'esident Wilson, 94235 Cachan cedex, France}

\begin{abstract}
The propagation of an adhesive crack through an anisotropic heterogeneous
interface is considered. Tuning the local toughness distribution function and
spatial correlation is numerically shown to induce a transition between weak to
strong pinning conditions. While the macroscopic effective toughness is given by
the mean local toughness in case of weak pinning, a systematic toughness
enhancement is observed for strong pinning (the critical point of the depinning
transition). A self-consistent approximation
is shown to account very accurately for this evolution, {\em without any free
parameter}.
\end{abstract}

\maketitle

{\em Introduction - } \Revision{While physicists studied the scaling
properties of crack\cite{Bouchaud-JPC97,Alava-AdP06} and developed
an analogy between crack front propagation and the dynamical phase
transition associated with the pinning/depinning of an elastic line
driven through a random
potential\cite{Schmittbuhl-PRL95,Daguier-PRL97,Ramanathan-PRL97a,
Ramanathan-PRL97b,Ramanathan-PRB98,Tanguy-PRE98,
Santucci-PRE07,DLV-PRL08,Laurson2010,Bonamy-PR11}, a parallel (and independent) effort was
made by mechanical engineers studying crack trapping by tough
particles\cite{GaoRice-JAM89,Bower-JMPS91,Mower-MM95} or the effect
of crack front deflection on the stress intensity factors (see
e.g. \cite{Lazarus-JMPS11} for a recent review).}

\Revision{Although the intimate link between surface energy of a material, and
the resisting ``force'', or toughness, opposing interfacial crack propagation
has been elucidated in the ideal cleavage case, the same concept remains to be
better understood in more common situations where solids are heterogeneous.
Generally, dissipative processes in the bulk of the solid (yet in the vicinity
of the crack surface) contribute to (or even dominates over) the thermodynamic
surface energy. This effect has been highlighted in recent studies
\cite{TvergaardIJSS2009,XiaPRL2012} showing how periodic modulations of elastic
or interface properties affect crack propagation and considerably enhance the
effective toughness of a given interface. Until
recently\cite{RVH-EJMA03,RH-IJF08,Demery2012}, however, the computation of such
an effective macroscopic toughness for random media has remained mostly
unexplored despite its great theoretical (critical point of the depinning
transition) and practical importance (optimized bonding). }

\begin{figure}[h!]
\centering
\includegraphics[width=0.99\columnwidth]{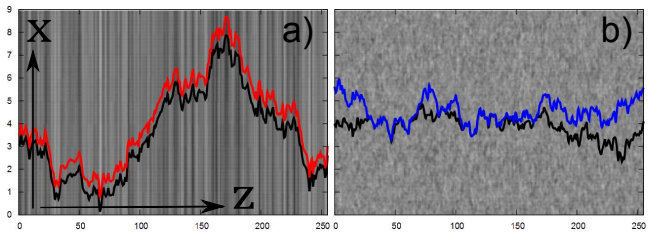}
\caption{\Revision{Snapshots of crack fronts for different toughness landscapes
  illustrating the transition from weak (a) to strong (b)
  pinning regimes \--- same distribution ($\sigma=0.125$), different correlation
  lengths: $\xi_x=50$ (a), $\xi_x=0.1$ (b).  Two
  fronts separated by a small time lag are represented. In weak
  pinning conditions, the distance between successive fronts is
  nearly constant, indicating weak fluctuations of the local velocity.
  In strong pinning conditions, motion is concentrated over small parts of the
  front, indicating a
  jerky dynamics.}}
\label{fig:front-vs-toughness}
\end{figure}

\Revision{When a crack propagates in a random solid heterogeneities may also trigger
  different dissipative phenomena, resulting in a toughness which
  cannot be reduced to the bare surface energy. Depending on the
  relative strength of the random potential and elasticity of the
  crack front, one usually distinguishes two
  generic situations\cite{Tanguy-EPJB04} (see
  Fig. \ref{fig:front-vs-toughness}):\\
  - In weak disorder conditions, the depinning front is only slightly
  perturbed and smoothly advances
  as a whole, with modest velocity fluctuations. Viscous dissipation
  can indeed be turned to arbitrary low values in quasi-static
  conditions, and hence only the bare {\em average} surface energy
  will be relevant for the macroscopic toughness. Disorder plays only
  a very minor role (e.g. for the geometry of the crack front).\\
  - In strong pinning conditions, the front advances intermittently, by
  series of localized micro-instabilities, the front roughness
  exhibits a non trivial scale free behavior. In contrast with the
  previous case, the local motion during a micro instability is no
  longer under the control of the experimentalist. The unbalance of
  elastic forces is compensated by local viscous friction until a
  new equilibrium configuration is reached. The external driving force
  does not interfere much with this local resolution of the
  disequilibrium. Yet, at a macroscopic scale, the accumulation of
  these micro-instabilities will contribute to a total energy
  dissipation that dresses the surface energy. A similar mechanism
  has long been proposed for solid
  friction~\cite{Heslot-PRE94,Caroli-EPJB98,Baumberger-AdvPhys06}, or
  plasticity~\cite{Truskinovsky-JMPS05,Truskinovsky-ARMA12}.}

\Revision{It is therefore crucial to quantify the onset of strong pinning and to
evaluate effective toughness in a quantitative fashion. Such is the purpose of
the present study. First a numerical model is introduced to account for the
crack front of an adhesive crack propagating in a random field of local
toughness.  This anisotropy-induced weak-to-strong pinning transition is shown
to severely affect the value of the depinning threshold, or macroscopic
toughness. Our results are shown to confirm early prediction about the effect of
anisotropy of toughness~\cite{RVH-EJMA03,RH-IJF08}. The velocity fluctuations along 
the front, through the participation ratio computation, are also shown to 
characterize the weak or strong pinning regimes.}

{\em Numerical model - } \Revision{We consider here an interfacial crack
front propagating in the $(z,x)$ plane in $x-$direction. The
location of the crack front at time $t$ is denoted $h(z,t)$. In the
framework of brittle fracture, propagation is ruled locally by the Griffith
criterion that compares a driving force, the Stress Intensity Factor
(SIF) $K$, that depends on the geometry and the external loading
with a threshold value $K_c$, the toughness, a material property.
The heterogeneity of the microscopic toughness is represented by a
random landscape $K_c(z,x)$ of mean $\overline{K_c}$ and standard
deviation $S_c$. It is assumed to be short-range correlated and its
correlation lengths are denoted $\xi_x$ and $\xi_z$ in the direction
of propagation and orthogonal to it respectively.}

\Revision{The microscopic toughness disorder induces a roughening of the crack
front, which in turn modifies the local value of the SIF along the front via a
long-range elastic restoration force\cite{GaoRice-JAM89}.  Neglecting inertial
effects, we consider in the following an over-damped dynamics. The local
(forward) velocity is here given by the positive part of the difference between
the local values of the SIF and the microscopic toughness\cite{Bonamy-PR11}:
$\mu \partial_th = {\cal R}\left(K-K_c\right)$ where $\mu$ stands for an
effective viscosity and $\cal{R}$ denotes the positive part. We study the
behavior of  the crack front at the verge of propagation from above, i.e. at a
vanishing velocity for $K\approx \overline{K_c}$. This justifies a first order
perturbative expansion around $\overline{K_c}$. The equation of evolution of the
crack front thus writes\cite{GaoRice-JAM89,Ramanathan-PRB98,Laurson2010}:}
\be
 \partial_t h(z,t) =
{\cal R}\left[  k_0(t)+ k_{el}(z,h(z,t)) - k_{c}(z,h(z,t))\right]
\label{eq:evol-reduced-1}
\ee
Here $k_c=K_c/\overline{K_c}$ is the reduced microscopic toughness
landscape of unit mean, of standard deviation
$\sigma=S_c/\overline{K_c}$ and of correlation lengths $\xi_x$ and
$\xi_z$.

The driving force $k_0(t)=K_0(t)/\overline{K_c}$ is the reduced average SIF
along the crack front.  \Revision{To account for the stiffness of the system
(specimen and loading device), the boundary condition is described by a slow and
steady loading rate such that on average the crack front velocity is set to
$v_0$.  A stiffness $e$ is introduced such that $k_0(t) = e\left[ v_0t -
\overline{h}(t) \right]$ where $\overline{h}(t)$ is the average position of the
front at time $t$ \cite{Bonamy2008}. The expression of the spatial modulation of
the reduced SIF $k_{el}$ as a function of the crack front geometry has been
obtained to first order in perturbation\cite{Rice-JAM85} and writes
  \be
  k_{el}(z,h(z))= \frac{1}{2\pi} \fint \frac{h(z')-h(z)}{(z-z')^2}dz'
  \ee
where $\fint$ stands for the principal value of the integral.}

Note that in Eq.(\ref{eq:evol-reduced-1}) the time scale has been set
so that the viscosity is scaled to unity.  Two parameters thus remain
that characterize the driving dynamics: the (reduced) stiffness $e$
and the (reduced) velocity $v_0$.  In the following only the
quasi-static limit $v_0\to 0^+$ is considered.

\begin{figure}[!h]
\centering
\includegraphics[width=0.99\columnwidth]{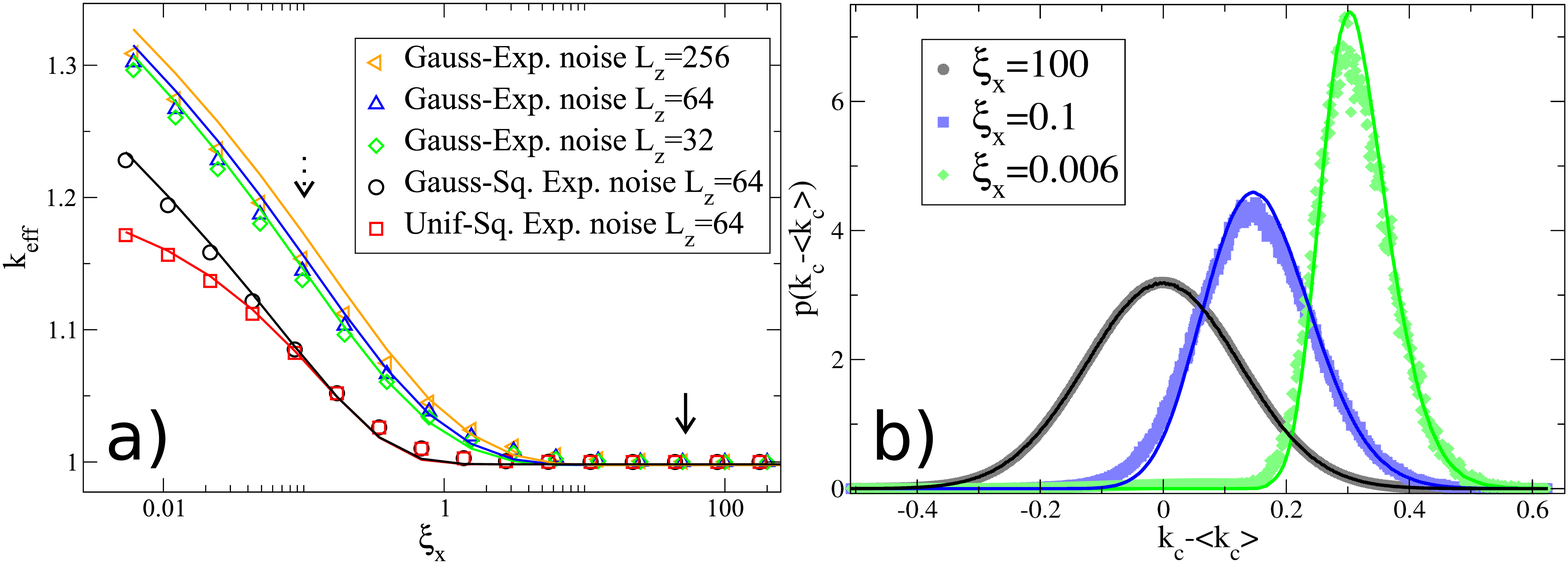}
\caption{\Revision{(a) Effective toughness $k_{eff}$ (for $\sigma=0.125$)
    vs correlation length $\xi_x$ for different 
    disorders types and system widths $L_z$. The continuous and dotted
    arrows indicate the weak and strong pinning regimes reported in
    Fig.~\ref{fig:front-vs-toughness}a and
    ~\ref{fig:front-vs-toughness}b respectively. (b) Toughness
    distribution along the crack front that propagates in an
    exponentially correlated Gaussian landscape of width $L_z=64$ for
    different $\xi_x$ ($\sigma=0.125$). Symbols indicate
    simulation data and lines theoretical predictions.}}
\label{fig:effective-toughness}
\end{figure}

In the following, random toughness fields of size $L_z\times L_x$ are considered with a unit lateral correlation length $\xi_z=1$ (the discretization length scale in the lateral direction) and a tunable correlation length $\xi_x$ in the direction of propagation. Three different types of random fields were considered: \Revision{Uniform-Squared Exponential (U-SE), Gaussian-Squared Exponential (G-SE) and Gaussian-Exponential (G-E) where the first term refers to the probability distribution function of $k_c$ and the second one qualifies its autocorrelation function $C(\Delta x)$ in the $x$ direction.
The U-SE and G-SE disorders consist of grids of $N_z(=L_z/\xi_z) \times N_x (=L_x/\xi_x)$ random numbers from uniform and Gaussian distributions respectively. The spacing between grids points in the $x$ direction follows a uniform distribution such that $C(\Delta x)=e^{-(\Delta x/\xi_{x})^2}$. The G-E landscape consists of realizations of an exponentially correlated Gaussian noise $C(\Delta x)=e^{-(\Delta x/\xi_{x})}$ computed} according to first order scheme\cite{Fox-PRA88} with an integration step $\Delta x=\xi_x/50$. In all cases, the continuous toughness landscape $k_c(z,x)$ is interpolated linearly between two grid points in the $x$ direction.

The standard deviation $\sigma$ is varied in the range $[0.125-1]$ and the
correlation length $\xi_x$ in the range $[0.006-800]$ while the reduced
stiffness is set to $e=1$. The chosen reduced velocity $v_0=\sigma/20$ was
verified to be small enough not to significantly influence the results. Periodic
boundary conditions along $z$ are considered. Integration of
Eq.~\ref{eq:evol-reduced-1} is performed according to an explicit mid-point
scheme. The time step $\delta t$ is chosen so that the maximum front increment
is less than on tenth of the noise discretization length. Starting from a flat
configuration, the crack front was first propagated over $\xi_x
(L_z/\xi_z)^{0.5}$ in order to reach a statistical steady state.

\Revision{In the spirit of a homogenization approach, the effective
toughness is defined as the one which would be measured at a macroscopic
scale. The (reduced) effective toughness $k_{eff}$ is thus measured as
the time (and ensemble) average of the (reduced) Stress Intensity Factor
$k_0(t)$ along propagation for a vanishing velocity: $k_{eff}=
\langle k_0(t) \rangle$. In practice $\keff$ is computed as the
mean value of the driving force minus the driving velocity $k_0-v_0$
along a propagation length equal to $L_x=1024\;\xi_x$.  This value is
finally averaged again over ten simulations (different statistical
samples).}

{\em Numerical results - } The weak-to-strong pinning transition induced by the
shortening of the toughness correlation length goes together with a spectacular
increase of the effective toughness (the depinning threshold).
{Fig.~\ref{fig:effective-toughness}a shows the dependence of $\keff$ on
different types of disorder. For large values of the correlation length $\xi_x$
we obtain $ \keff= 1$ i.e. the effective toughness equals the mean of the
microscopic disorder $\langle k_c\rangle$.} However for low values of $\xi_x$,
$\keff$ departs from $\langle k_c\rangle=1$ and reaches significantly higher
values that clearly depend on the type of distribution and correlation of the
microscopic disorder. Indeed the effective toughness is interpreted here as the
threshold of a (dynamic) phase transition and as such is expected to depend on
the microscopic details. We also note a slight but clear dependence on the
system size: the larger the system, the larger $\keff$.  This enhancement is
also reflected by changes in the toughness distribution (weighted by time)
visited by the crack front as illustrated in Fig.~\ref{fig:effective-toughness}.
{This distribution is clearly biased towards higher toughness values as $\xi_x$
decreases.}
\begin{figure}[h]
\includegraphics[width=0.95\columnwidth]{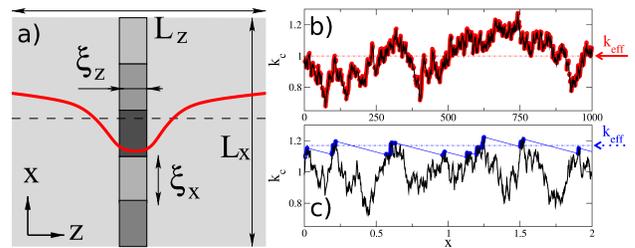}
\caption{\Revision{(a) A band of random toughness in a homogeneous landscape
  traps the crack front; the lower the line stiffness, the larger the
  front deflection. (b) and (c) Graphical solutions of the implicit
  equation $\Delta k_{el}(x)= \Delta k_{c}(x)$ which determines the equilibrium
  position of the crack front for the same disorder parameters reported in
  Fig.~\ref{fig:front-vs-toughness}a and ~\ref{fig:front-vs-toughness}b.
  The stable positions in the band are highlighted in red and blue respectively.
 For large correlation length (b), the trajectory almost follows the
  toughness in the band (weak pinning). For small correlation length (c), the
  equilibrium positions are skewed toward higher
  toughness (strong pinning). The slope of the fine lines between the stable
  portions corresponds to the stiffness $S$ of the crack front.}}
\label{fig:self-consistent}
\end{figure}

{\em Interpretation: a self-consistent approach - } This evolution can be
rationalized in the framework of a self-consistent approximation proposed in
Ref.~\cite{RVH-EJMA03}, and never checked quantitatively. The top panel of
Fig.~\ref{fig:self-consistent} shows a toughness landscape consisting of a
unique band of fluctuating toughness $k_c(x)$ in an otherwise homogeneous medium
of toughness $k_0$. Because of the elastic coupling, the crack front undergoes
a deflection $\Delta h=\overline{h}-h(0)$ proportional to the toughness contrast
$\Delta k=k_c-k_0$. The associated line stiffness $S=\Delta k/\Delta h$ can be
analytically computed from Rice formula\cite{RVH-EJMA03,Chopin-PRL11} and is
shown to scale as $S \propto 1/\xi_z\log(L_z/\xi_z)$.  An effective medium
approximation, in the spirit of the self-consistent approximation, consists in
evaluating $\keff$ as equal to the value of $k_0$ such that the average
deflection of the front taken over the ensemble of successive {\em stable}
positions of the front is zero.

{\em From weak to strong pinning - } The simplicity of the one-dimensional
picture of Fig.~\ref{fig:self-consistent} allows one to define a criterion that
determines the pinning regime.  In absence of driving force, an equilibrium
configuration of the front at position $h(z=0,t)=x$ in the band is obtained when
the elastic restoring force, $\Delta k_{el}(x)=S\Delta h(x)$, balances the
toughness contrast $\Delta k_c(x)$. Depending on the respective amplitude of the
line stiffness and the local toughness gradient, one can obtain for this
implicit equation either a unique solution or multiple solutions for the front
position $x$. A simple criterion for the transition from weak to strong pinning
can thus be drawn from the onset of multistability. Under these conditions,
strong pinning is obtained when
\be
\frac{\sigma}{S\xi_x} > 1.
\label{eq:criterion}
\ee
As schematically shown in Fig.~\ref{fig:self-consistent}b different sets
of stable positions along the band are generated as a function of the line
stiffness. For large $S\xi_x/\sigma$ the stable trajectory closely follows the
toughness of the band as expected from the weak pinning regime. Conversely for
lower $S\xi_x/\sigma$ the crack front will only visit a subset of high toughness
values that characterize the strong pinning regime. As a consequence, the
distribution of toughness at stable positions is skewed toward higher values.

\begin{figure}[h]
  \centering
\includegraphics[width=0.9\columnwidth]{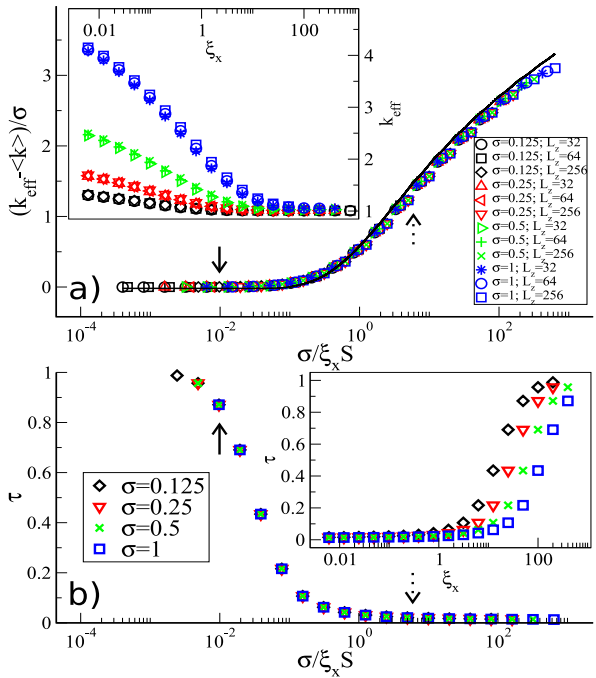}
\caption{\Revision{(a) Effective toughness $k_{eff}$ as a function of the
    correlation length $\xi_x$ for different landscape toughness
    standard deviations $\sigma$ and system widths $L_z$.  The larger
    $\sigma$ and/or the shorter $\xi_x$, the larger $k_{eff}$. Symbols
    and lines correspond to simulation data and theoretical
    predictions respectively. (b) Participation ratio $\tau$ along a
    crack front of width $L_z=256$. The larger $\sigma$ and/or the
    shorter $\xi_x$, the lower $\tau$. In both figures, disorder type
    is GE. Original simulation data are reported in the insets while the main
    panels correspond to the rescaled quantities following
    Eq.~\ref{eq:criterion}.  The
    continuous and dotted arrows indicate the weak and strong pinning
    regimes reported in Fig.~\ref{fig:front-vs-toughness}a and
    ~\ref{fig:front-vs-toughness}b respectively.}}
\label{fig:collapse}
\end{figure}

Direct numerical integrations of the effective medium model have been performed.
$S$ is first determined numerically to account for the discreteness of the
simulated crack front. The reduced toughness distributions and their means are
then generated from one dimensional trajectories as exemplified in
Fig.~\ref{fig:self-consistent}a and b. The comparison between the
self-consistent approximation and the crack front simulations reported in
Fig.~\ref{fig:effective-toughness} and in Fig.~\ref{fig:collapse} shows a
remarkable agreement. $\keff$ is very accurately reproduced as a function of
$\xi_x$, $L_z$, $\sigma$ and the different disorder types. Note that the model
not only accounts for the effective toughness variations but also for the
visited toughness distributions {\em without any free parameter}.

{\em Scaling analysis - } \Revision{One can show that only a combination of
scaled parameters really contributes to the observed results. If the toughness
$k$, and distance $x$ along the propagation direction, are transformed into $k
\rightarrow (k-\langle k_c\rangle)/\sigma$ and $x \rightarrow x/\xi_x$ then the
elasticity of the crack front line becomes $S\xi_x/\sigma$.
Fig.~\ref{fig:collapse}a shows the centered and reduced effective toughness
against the depinning control parameter $\sigma/S\xi_x$. All numerical results
collapse onto a single master curve that captures the transition. For small
values of the control parameter $\sigma/S\xi_x$ the weak pinning regime holds,
$\keff=1$, whereas, for large values $\sigma/S\xi_x$, $\keff$ significantly
increases (strong pinning).} Note that the system size dependence is here
captured in the line stiffness parameter which decreases as the logarithm of the
crack front length. This has no effect in the weak-pinning regime, but justifies
a rather counter-intuitive result in the framework of brittle fracture: the
larger the system, the smaller the line stiffness and hence the larger the
effective toughness.

Details of the local disorder have a dramatic effect on the pinning conditions.
A crack front can encounter weak or strong pinning depending on its propagation
direction. In the same spirit, if the toughness landscape is non-symmetrical,
the effective toughness in a given direction will not be the same as in the
reverse one. Our approach thus offers a natural interpretation to the recent
results presented in Ref.~\cite{XiaPRL2012}.

\Revision{{\em Crack front dynamics - } As mentioned above, the
  weak-to-strong pinning transition is also characterized by the
  emergence of intermittence and localization of the propagation. To
  quantify the localization degree along the front,
  the participation ratio~\cite{Tanguy-EPJB04} is computed
\be
\tau=\left\langle \frac{(\sum_{i=1}^{N_z}\delta h(z_i)^2)^2}{N_z\sum_{i=1}^{N_z}\delta h(z_i)^4}\right\rangle ,
\label{eq:PartRatio}
\ee
where $\delta h(z_i)$ is the local velocity of the $i$th site and $N_z$ the
total number of site along the crack front. The brackets denote time average.
This scalar parameter measures the relative number of sites involved during
motion. In the weak pinning case where all sites move, $\tau=1$. In the strong
pinning limit where only one site moves, $\tau=1/N_z$ is expected.  The
numerical results reported in Fig.~\ref{fig:collapse}b are fully consistent with
this picture: the lower the correlation length $\xi_x$ and/or the larger the
standard deviation $\sigma$ of the toughness disorder, the lower the
participation ratio, i.e. the wider the velocity distribution. As shown in
Fig.~\ref{fig:collapse}b, the participation ratio data collapses onto a master
curve describing the transition from weak to strong pinning for different
disorder strengths and correlation lengths, validating again the relevance of
the scaling parameter.}

{\em Conclusion - } The depinning of a crack front has been shown to be strongly
dependent on the spatial correlation of the disordered landscape through which
it propagates. The transition between weak and strong pinning is well captured
by a simple criterion built on the toughness gradient in the direction of
propagation and the line stiffness. A simple self-consistent approximation
very accurately describes the progressive departure of the depinning threshold
from its weak pinning value (the mean value of the disorder) to the higher
values measured for strong pinning. This scheme accurately captures the
dependence of the depinning threshold to the finer microscopic details
(statistical distribution of toughness and  spatial correlation function).  Let
us emphasize that such a result may not have been expected from the fact that
strong pinning involves micro-instabilities and collective phenomena
(avalanches, ...) due to the underlying dynamical phase transition.

Bearing in mind that the enhancement of the effective toughness does not
originate from the initial landscape toughness distribution only but also from
its spatial correlation, our results open a promising route to design
anisotropic and tough interfaces.

\begin{acknowledgments}
S.P. acknowledges the support of ANR project MePhyStaR.
\end{acknowledgments}

\bibliography{PVR}

\end{document}